\documentclass[
	preprint,
	aps,
	pra,
	english
	]{revtex4-1}

\usepackage[utf8]{inputenc}
\usepackage{amsmath}
\usepackage{fixmath}
\usepackage{dsfont}
\usepackage{graphicx}
\usepackage{hyperref}
\usepackage{xcolor}
  \newcommand*{\unity}{\mathds{1}}
  \newcommand{\conj}[1]{#1^*}
  \renewcommand*\d[1]{\mathrm{d} #1\,}
  \newcommand{\ii}{\mathrm{i}}
  \newcommand*{\per}{\operatorname{perm}}
  \newcommand{\abs}[1]{\left\lvert{#1}\right\rvert}
  \newcommand{\ee}[1]{\mathrm{e}^{#1}}
  \newcommand*{\defeq}{\mathrel{\vcenter{\baselineskip0.5ex \lineskiplimit0pt
                     \hbox{\scriptsize.}\hbox{\scriptsize.}}}%
                     =}
  \renewcommand{\Omega}{\varOmega}
  \renewcommand*\vec[1]{\mathbold{#1}}
  
  \newcommand{\ket}[1]{\vert #1 \rangle}

  \newcommand{\matrixel}[3]{\langle #1 \vert #2 \vert #3 \rangle}

  \newcommand*\setS{\mathcal{S}}
  \newcommand*\setD{\mathcal{D}}

  \newcommand*\Mports{M}
  \newcommand*\Ndets{N}
  
  \newcommand*\permutd{\delta}
  \newcommand*\detd{d}
  \newcommand*\sources{s}
  
  \newcommand*\SymmGroup[1]{\Sigma_{#1}}
  \newcommand*\PathGroup{\Omega}
  \newcommand*\BSPindex{\text{av}}

  \newcommand*\TotalProb[1]{P_{\BSPindex}(\setD;\setS)}
  \newcommand*\ProdMatrix[2]{\mathcal{A}^{(\setD,\setS)}_{#1}}
  \newcommand*\OverlapFrequency[2]{g(#1,#2)}
  \newcommand*\OverlapFactor[1]{f_{#1}(\setS)}
  \newcommand*\TotalProbIntervalSingle[1]{P\left( \tdettintervalSet, \setD; \setS \right)}

  \newcommand*{\Umatrix}[1]{\mathcal{U}^{(\setD,\setS)}_{#1}}

  \newcommand*\Uto[2]{\mathcal{U}_{#2,#1}}
  \newcommand*\Utoconj[2]{\conj{\mathcal{U}}_{#2,#1}}

  \newcommand*\tdet[1]{t_{#1}}

  \newcommand*\tinterval[1]{\Delta \tdet{#1}}
  \newcommand*\tdettintervalSet{ \left\{ \tdet{\detd}, \tinterval{\detd} \right\}}

\begin{document}


\title{Boson sampling with non-identical single photons}

  \author{Vincenzo Tamma}
  \email{vincenzo.tamma@uni-ulm.de}
  \author{Simon Laibacher}
  \affiliation{Institut f\"{u}r Quantenphysik and Center for Integrated Quantum
  Science and Technology (IQ\textsuperscript{ST}), Universit\"at Ulm, D-89069 Ulm, Germany}

\begin{abstract}
The boson sampling problem has triggered a lot of interest in the
scientific community because of its potential of demonstrating the
computational power of quantum interference without the need of non-linear
processes. However, the intractability of such a problem with any classical
device relies on the realization of single photons approximately identical
in their spectra. In this paper we discuss the physics of boson
sampling with non-identical single photon sources, which is strongly
relevant in view of scalable experimental realizations and triggers
fascinating questions in complexity theory.
\end{abstract}

\maketitle

\section{Introduction}

The realization of schemes in quantum information processing (QIP) has raised great interest in the scientific community because of their potential to outperform classical protocols in terms of efficiency. Quantum interference and quantum correlations \cite{TammaLaibacher2015,LaibacherTammaComplexity} are two fundamental phenomena of quantum mechanics at the heart of the computational speed-up and metrologic sensitivity attainable in QIP. 

Recently, a strong experimental effort \cite{Broome2013,Crespi2013,Spring2013,tillmann2013experimental,Tillmann2014} has been devoted towards the solution of the boson sampling problem (BSP) \cite{aaronson2011computational,Ralph2013,Franson2013,Gard2014}, which consists of sampling from all the possible detections of $N$ single input bosons, e.g. photons, at $N$ output ports of a linear {$2M$-port} interferometer, with $M \gg N$ input/output ports. In particular, the BSP has an enormous potential in quantum computation  thanks to its connection with the computation of permanents of random matrices \cite{Minc1984}, in general more difficult than the factorization of large integers \cite{Valiant1979}. Moreover, any interferometric network directed towards the implementation of the BSP has the advantage of not requiring entangled sources. However, in its current formulation, linear interferometric networks able to physically implement the BSP encounter the experimental challenge of producing approximately identical photons.

In \cite{TammaLaibacher2015} we demonstrated that, even for non-identical photons,  the problem of multi-boson correlation sampling, i.e.~sampling by time and polarization resolving correlation measurements from the output probability distribution of a random linear interferometer, is, in general, intractable with a classical computer.

Here, we focus on the original boson sampling problem \cite{aaronson2011computational} based on measurements not able to resolve the detection times and the detected polarizations for single input photons in arbitrary pure states. 

 \section{Interferometer description}

\begin{figure}
	\begin{center}
		\includegraphics{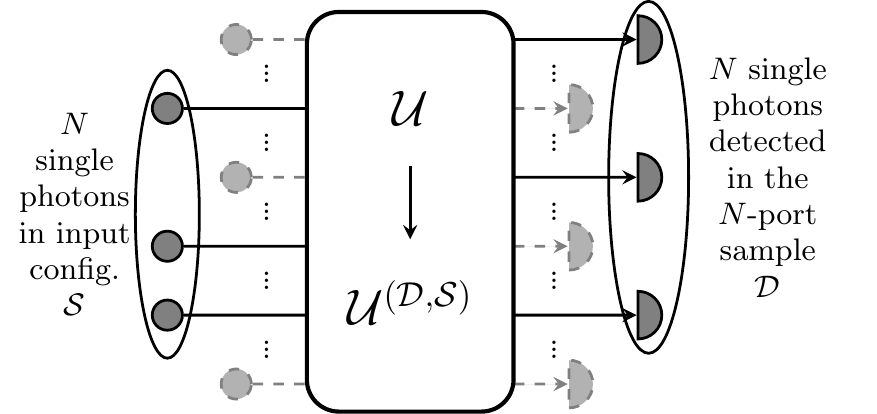}
		\caption{
			General boson sampling interferometer. $N$ single bosons are injected into an $N$-port subset $\mathcal{S}$ of the $M$ input ports of a linear interferometer. They can be detected at the output in any possible sample $\mathcal{D}$ containing $N$ of the $M$ output ports. For each output port sample $\mathcal{D}$ and given input configuration $\mathcal{S}$, the evolution through the interferometer is fully described by an $N\times N$ submatrix $\mathcal{U}^{(\mathcal{D},\mathcal{S})}$ of the original $M\times M$ interferometer random matrix $\mathcal{U}$.}
		\label{fig:InterferometerSetup}
	\end{center}
\end{figure}
We consider the linear interferometer in Fig.~\ref{fig:InterferometerSetup}. 
For a set $\mathcal{S}$ of occupied input ports, the input state 
\begin{align}
	\ket{\mathcal{S}} \defeq \bigotimes_{s\in \mathcal{S}}
	\ket{1[\vec{\xi}_{s}]}_{s}
	\bigotimes_{s \notin \mathcal{S}}
	\ket{0}_{s},
	\label{eqn:StateDefinition}
\end{align}
is defined by $N$ single-photon pure states 
\begin{align}
	\ket{1[\vec{\xi}_{s}]}_{s} \defeq \sum_{\lambda=1,2} \int_{0}^{\infty} \d{\omega} \left( \vec{e}_{\lambda}\cdot \vec{\xi}_{s}(\omega) \right) \hat{a}^{\dagger}_{s,\lambda} (\omega) \ket{0}_{s},
	\label{eqn:SinglePhotonState}
\end{align}
with the creation operator $\hat{a}^{\dagger}_{s,\lambda}(\omega)$ for
the frequency mode $\omega$ and the polarization mode $\lambda$ \cite{Loudon2000}. The magnitude, phase and direction of the complex spectral distribution $\vec{\xi}_{s}(\omega)$  (normalized as $\int
\d{\omega} \abs{\vec{\xi}_{s}(\omega)}^2=1$) correspond to the frequency spectrum, the time of emission, and the polarization of the photon, respectively.

After the evolution in the interferometer, in the condition of equal propagation times $\Delta t$ along any possible path, an $N$-photon detection can occur in any $N$-port sample $\mathcal{D}$. The rate of an $N$-fold detection event at detection times $\{t_d\}_{d\in \mathcal{D}}$ and in the polarizations $\{\vec{p}_{d}\}_{d\in \mathcal{D}}$ is given by the $N$th-order Glauber correlation function \cite{Glauber2007}
\begin{align}
	G^{(\mathcal{D},\mathcal{S})}_{\{t_d,\vec{p}_d\}} \defeq
	\matrixel{\mathcal{S}}{\prod_{d\in \mathcal{D}} \Big(\vec{p}^{*}_d \cdot \hat{\vec{E}}^{(-)}_d(t_d) \Big) \Big( \vec{p}_d \cdot \hat{\vec{E}}^{(+)}_d (t_d) \Big) }{\mathcal{S}}, 
	\label{eqn:ProbabilityRate}
\end{align}
where $\vec{p}_d\cdot \hat{\vec{E}}^{(+)}_d (t_d)$ are the $\vec{p}_d$-polarized components of the electric field operators 
\begin{align}
	\hat{\vec{E}}^{(+)}_d(t_d) = \sum_{s\in \mathcal{S}}
	\mathcal{U}_{d,s} \hat{\vec{E}}^{(+)}_s (t_d-\Delta t),
	\label{eqn:ExpansionAout}
\end{align}
with $d\in \mathcal{D}$, written in terms of the operators $\hat{\vec{E}}^{(+)}_{s} (t_d-\Delta t)$ at the input ports $s\in \mathcal{S}$, and the $N \times N$ submatrix
\begin{align}
	\mathcal{U}^{(\mathcal{D},\mathcal{S})} \defeq [ \mathcal{U}_{d,s} ]_{\substack{d\in \mathcal{D} \\ s\in \mathcal{S}}}
	\label{eqn:SubU}
\end{align}
of the {Haar} random unitary $M\times M$ matrix $\mathcal{U}$ describing the interferometer 
(requiring $\mathcal{U}$ to be a Haar random matrix is important in the discussion of the complexity of boson sampling but the detection probabilities derived in this article hold for general $\mathcal{U}$). 
$\mathcal{U}^{(\mathcal{D},\mathcal{S})}$ is obtained by first using the columns $s\in \mathcal{S}$ of $\mathcal{U}$ to form a $M\times N$ matrix and then only retaining the rows $d\in \mathcal{D}$, repeating rows as many times as the corresponding output port index is part of $\mathcal{D}$.
We refer to \cite{TammaLaibacher2015} for  the full description of the physics of multi-photon correlation interferometry in arbitrary linear networks arising from the correlation function in Eq. (\ref{eqn:ProbabilityRate}), which describes all the interfering $N$-photon detection amplitudes for an $N$-photon detection event \cite{Glauber2007}.

\section{Boson sampling based on ``non-resolving'' measurements}

Here, we discuss only the case of measurements which do not resolve the detection times and polarizations, resulting in an average over these degrees of freedom. This is the type of measurement considered in the initial formulation of the boson sampling problem introduced in \cite{aaronson2011computational}. In this case, we obtain the probability \cite{TammaLaibacher2015}
\begin{align}
	P_{\text{av}}(\mathcal{D};\mathcal{S}) \defeq
	\sum_{\{\vec{p}_d\} \in \{ \vec{e}_1, \vec{e}_2 \}^{N}} \int_{-\infty}^{\infty} \Big(
	\prod_{d\in \mathcal{D}} \d{t_d} \Big) G^{(\mathcal{D},\mathcal{S})}_{\{t_d,\vec{p}_d\}}
	\label{eqn:IntegratedNotExplicit}
\end{align}
to detect the $N$ photons injected in the ports $\mathcal{S}$ at the output
ports $\mathcal{D}$, where $\left\{ \vec{e}_1,\vec{e}_2 \right\}$ is an arbitrary polarization basis.

It is useful to define the \textit{interference-type matrices}
\begin{align}
	\mathcal{A}_{\rho}^{(\mathcal{D},\mathcal{S})}\defeq
	[ \mathcal{U}_{d,s}^{*} \mathcal{U}_{d,\rho(s)} ]_{\substack{d\in \mathcal{D} \\
	s\in \mathcal{S}}}.
	\label{A}
\end{align}
and the \textit{indistinguishability weights}
\begin{align}
	f_{\rho}(\mathcal{S}) &\defeq
	\prod_{s\in \mathcal{S}} g(s,\rho(s)),
	\label{eqn:DistinguihsabilityFactor}
\end{align}
with a permutation $\rho$ from the symmetric group $\Sigma_N$, and with the two-photon indistinguishability factors
\begin{align}
g(s,s') = \int_{-\infty}^{\infty} \d{\omega}
	\vec{\xi}_s(\omega)\cdot \vec{\xi}_{s'}(\omega). 
	\label{eqn:g}
\end{align}
In general, (in-)distinguishability of photons must be defined at the detectors since the ability to distinguish the photons strongly depends on the detection scheme applied and on the time delays which the photons undergo during the propagation through the interferometer. In the setup considered here, where the propagation times are all equal and the detectors average over the detection times and polarizations, the indistinguishability of the input photons at the detectors is determined solely by the overlap of their complex spectral distributions $\vec{\xi}_s(\omega)$.

As shown in \cite{TammaLaibacher2015} the probability of an $N$-fold detection in the sample $\mathcal{D}$ can be expressed, in the narrow-bandwidth approximation, as 
\begin{align}
	P_{\text{av}}(\mathcal{D};\mathcal{S})
	= \sum_{\rho \in \Sigma_N} f_{\rho}(\mathcal{S}) \per \mathcal{A}_{\rho}^{(\mathcal{D},\mathcal{S})}
	= \sum_{\rho\in \SymmGroup{\Ndets}}
		\OverlapFactor{\rho} \sum_{\permutd\in\PathGroup} \left[
					\prod_{\sources\in\setS} \Utoconj{\sources}{\permutd(\sources)}
							\prod_{\sources\in\setS}\Uto{\rho(\sources)}{\permutd(\sources)} 
								 \right],
		\label{eqn:IntegratedGeneral} 
\end{align}
where $\Omega$ is the group of bijective functions $\delta$ between the sets $\mathcal{S}$ and $\mathcal{D}$.

The probability $\TotalProb{\BSPindex}$ in
Eq.~\eqref{eqn:IntegratedGeneral} associated with the detection of $\Ndets$ photons in the
$\Ndets$-port sample $\setD$ comprises $\Ndets!$ contributions for each
permutation $\rho\in\SymmGroup{\Ndets}$. 
Each contribution contains all $\Ndets!$ terms $\prod_{\sources\in\setS}
\Utoconj{\sources}{\permutd(\sources)} \Uto{\rho(\sources)}{\permutd(\sources)}$
arising from the interference of the interferometer-dependent multi-photon
amplitudes $\prod_{\sources\in\setS}\Uto{\sources}{\permutd(\sources)}$, with
$\permutd\in\PathGroup$, in the condition that the $\Ndets$ source pairs
$\left\{(\sources,\rho(\sources))\right\}_{\sources\in\setS}$ for each cross term are fixed by a given permutation $\rho$. Moreover, each weight
$\OverlapFactor{\rho}$
describes the degree of pairwise indistinguishability for the set of source pairs $\left\{(\sources,\rho(\sources))\right\}_{\sources\in\setS}$. 

In the trivial case $\rho=\unity$, each photon is paired with
itself and the indistinguishability weight is $\OverlapFactor{\rho} = 1$. Yet, the
indistinguishability weight for $\rho \neq \unity$ can have arbitrary values depending on the
overlap of the spectral distributions $\vec{\xi}_{\sources}(\omega)$ and $\vec{\xi}_{\rho(\sources)}(\omega)$ of each pair of photons. Thereby, these weights describe how the degree of distinguishability of the photons at
the detectors affects the degree of interference between the multi-photon
quantum paths for each pairwise configuration $\left\{ (\sources,
\rho(\sources)) \right\}_{\sources\in\setS}$.

\begin{figure}
	\begin{center}
		\includegraphics[scale=0.98]{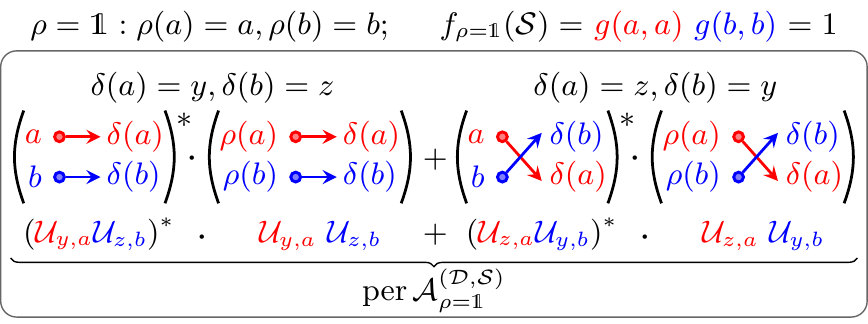} 
		\includegraphics[scale=0.98]{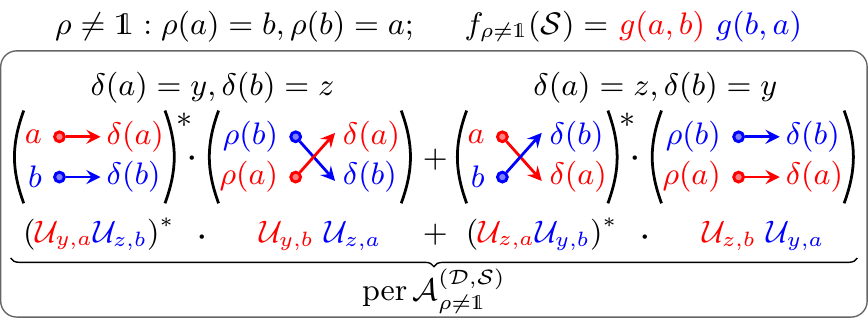} 
		\caption{Physical illustration of the permanent terms in
			Eq.~\eqref{eqn:IntegratedGeneral} for $\Ndets=2$ ($\setS=\left\{ a,b
			\right\}$,$\setD=\left\{ y,z \right\}$) in the two possible cases
			$\rho=\unity$ (indistinguishability weight
			$\OverlapFactor{\rho=\unity}=1$) and $\rho\neq\unity$
			($\OverlapFactor{\rho\neq\unity}\neq 1$ for incomplete overlap of the
			photon spectra). In both cases,
			$\Ndets=2$ pairs of sources $\left\{ (\sources,\rho(\sources))
			\right\}_{\sources=a,b}$, each indicated by a separate color, are
			coupled in the $\Ndets!=2$ possible interference terms defined by the
			two ways of connecting the two pairs to the two detectors $y$ and
			$z$. These possibilities sum up to the permanents $\per
						\mathcal{A}^{(\setD,\setS)}_{\rho} $.}
		\label{fig:InterferenceTerms}
	\end{center}
\end{figure}

In Fig.~\ref{fig:InterferenceTerms} we illustrate as an example the case of
$\Ndets=2$. In this case the probability in Eq.~\eqref{eqn:IntegratedGeneral} becomes
\begin{align}
\TotalProb{\BSPindex} &= 
\per \mathcal{A}_{\rho=\unity}^{(\mathcal{D},\mathcal{S})} + \abs{g(a,b)}^2 \per \mathcal{A}_{\rho\neq \unity}^{(\mathcal{D},\mathcal{S})} \\
 &= \abs{\Uto{a}{y}\Uto{b}{z}}^2 + \abs{\Uto{b}{y}\Uto{a}{z}}^2 + \abs{\OverlapFrequency{a}{b}}^2 2 \Re\left( \Utoconj{a}{y}\Utoconj{b}{z} \Uto{b}{y}\Uto{a}{z} \right).
\end{align}						
If the interferometer reduces to a simple balanced beam splitter described by the unitary matrix
\begin{align}
	\mathcal{U}^{(\mathcal{D},\mathcal{S})} = \frac{1}{\sqrt{2}} \left( \begin{array}{cc} 1 & \ii \\ \ii & 1 \end{array} \right),
\end{align}
we obtain the expression 
\begin{align}
	\TotalProb{\BSPindex} = \frac{1}{2}\left( 1 - \abs{g(a,b)}^2 \right).
	\label{eqn:TwoPhotDipGeneral}
\end{align}
If the two photons have equal polarizations and Gaussian spectral distributions
\begin{align}
	\xi_s(\omega) = \frac{1}{(2\pi \Delta\omega^2)^{1/4}} \exp\left( -\frac{(\omega-\omega_0)^2}{4\Delta\omega^2} + \ii \omega t_{0,s} \right),
\end{align}
with $s = a,b$, that only differ in the central times $t_{0,s}$ of the photon pulses (equal frequency bandwidth $\Delta \omega$ and central frequency $\omega_0$), the two-photon indistinguishability factor $g(a,b)$ becomes a Gaussian in the initial time difference $t_{0,b} - t_{0,a}$ and Eq.~\eqref{eqn:TwoPhotDipGeneral} reduces to
\begin{align}
	\TotalProb{\BSPindex} = \frac{1}{2} \left( 1 - \ee{ - (t_{0,b} - t_{0,a})^2 \Delta\omega^2/2} \right),
\end{align}
describing the well known two-photon interference ``dip'' \cite{Hong1987,Shih1988}.

In general, the probability   $\TotalProb{\BSPindex}$ in
Eq.~\eqref{eqn:IntegratedGeneral} generalizes the description of multi-photon interference from the two-photon interference at a beam splitter to the $\Ndets$-photon interference emerging in a linear interferometer with $2\Mports \geq 2\Ndets$ ports.
This result reduces to the one obtained for $\Ndets=3$ in
Refs.~\cite{Tan2013,deGuise2014}  if we assume Gaussian spectra $\vec{\xi}_{\sources}(\omega)$ corresponding to equally polarized photons emitted at different times.
Differently from \cite{Tan2013,deGuise2014},
our result describes multi-photon interference of arbitrary order $\Ndets$. Moreover, the approach in \cite{Tan2013,deGuise2014} relies on the use of immanants \cite{littlewood1934group}, while our description depends on the sum of  matrix permanents weighted by the $\Ndets$-photon
indistinguishability weights in Eq.~\eqref{eqn:DistinguihsabilityFactor}. 
A comparison of both approaches may lead to a deeper understanding of the complexity of boson sampling with non-identical photons.

\subsection*{Limiting cases of multi-boson distinguishability}

We consider now three limiting scenarios of multi-photon distinguishability:

1) \textit{Complete $\Ndets$-boson distinguishability :}
In this case, there is no pairwise overlap of the spectral distributions at the
		detectors ($\OverlapFrequency{\sources}{\sources'}\approx
		\delta_{\sources,\sources'}\ \forall \sources,\sources'\in\setS$) and the
		indistinguishability weights $\OverlapFactor{\rho} \approx 0 \ \forall
		\rho\neq\unity$. Thus the probability in
		Eq.~\eqref{eqn:IntegratedGeneral} is given by the completely
		incoherent superposition
\begin{align*}
	\TotalProb{\BSPindex} &\approx \per \ProdMatrix{\rho=\unity}{} 
\end{align*}
with the non-negative matrix
\begin{align*}
	\ProdMatrix{\rho=\unity}{} = \left[ \abs{\Uto{\sources}{\detd}}^2
	\right]_{\substack{\detd\in\setD \\ \sources\in\setS}},
\end{align*}
whose permanent can be efficiently estimated \cite{Jerrum2004}. Because of the complete distinguishability of the $\Ndets$ photons at each
detector, all multi-photon quantum paths are distinguishable and no multi-photon interference occurs; the problem is computationally easy.

2) \textit{Complete $\Ndets$-boson indistinguishability:}
	If all photon spectra fully overlap pairwisely, the 
indistinguishability weights $\OverlapFactor{\rho}=1 \ \forall \rho$.
Accordingly, Eq.~\eqref{eqn:IntegratedGeneral} reduces to
\begin{align}
	\TotalProb{\BSPindex} & \approx 
	\sum_{\rho \in \SymmGroup{\Ndets}}^{} \per \ProdMatrix{\rho}{}
	= \abs{\per \Umatrix{}}^2,
	\label{eqn:IdenticalPhots}
\end{align}
corresponding to the full interference of the $\Ndets!$ interferometric
amplitudes associated with all possible multi-photon quantum paths. For each possible sample $\mathcal{D}$, such multi-photon interference leads to an $N$-photon detection probability depending on the permanent of random complex matrices, leading to the computational complexity of the boson sampling problem \cite{aaronson2011computational}.
Here, all $\Ndets!$ quantum paths from the $\Ndets$ sources in $\setS$ to the
$\Ndets$ detectors in $\setD$ superpose coherently, giving rise to the permanent of the
interferometer submatrix $\Umatrix{}$. Each of the quantum paths is associated
with a term in the permanent. The modulus squared of the resulting coherent superposition
expresses the full interference of all $\Ndets!$ quantum paths.
	
3) \textit{Complete $\Ndets'$-boson indistinguishability and $(\Ndets-\Ndets')$-boson distinguishability:}
We consider the case where $N'$ photons in the input port set $\Phi$ are fully indistinguishable from each other ($g(s,s')=1 \ \forall s,s'\in \Phi$) while the remaining $N-N'$ photons are fully distinguishable from all other photons ($g(s,s') = \delta_{s,s'} \ \text{if } s\notin\Phi \text{ or } s'\notin\Phi$). Then,
for each possible set $\Theta\subset\setD$ of
$\Ndets'$ detectors, multi-photon interference occurs only among the $\Ndets'$ photons
in the subset $\Phi\subset\setS$ of input ports.
Each of the remaining $(\Ndets-\Ndets')$
photons in the input ports $\sources\in\setS\setminus\Phi$ does not interfere with any other photon.
Thereby, the probability in Eq.~\eqref{eqn:IntegratedGeneral} becomes
\begin{align}
	\TotalProb{\BSPindex} &\approx 
	\sum_{\Theta}^{} \per \mathcal{A}_{\rho=\unity}^{(\setD\setminus\Theta,\setS\setminus\Phi)} \abs{\per \mathcal{U}^{(\Theta,\Phi)}}^2.
	\label{NN'}
\end{align}

In this case, the probability associated with each possible sample $\mathcal{D}$ depends on a weighted sum of $N!/(N'!(N-N')!)$ permanents of complex random $N'\times N'$ matrices, with $N'<N$. Already for $N'\gtrsim 30$ the computation of any of these permanents is not tractable with a classical computer. Thereby, one may ask: what is the complexity of the weighted sum of such permanents describing the detection probability in Eq. (\ref{NN'})? As far as we know, complexity theory has not yet an answer to such a question. 

\section{Discussion}

We summarize here our discussion about the physics and complexity of boson sampling for input single photons with non-identical spectra. The sampling process \cite{aaronson2011computational} at the detectors {averages} over the arrival times and the polarizations of the detected photons. This leads to the probability in Eq. (\ref{eqn:IntegratedGeneral}) of detecting $N$ single photons in a sample $\mathcal{D}$ of the interferometer output ports, independently of time and polarization. 

This result applies to arbitrary interferometric configurations with $N\leq M$ and generalizes the two-photon interference at a beam splitter to arbitrary $N$-order interference ``landscapes''. Moreover, it is relevant in the description of multiphoton interferometry with arbitrary Fock states
\cite{TammaLaibacher1} and other different sources \cite{TammaLaibacher2,Lund2014,Rahimi2015,Rohde2015,Seshadreesan2015}, as well as in the analysis of sampling of bosonic qubits \cite{Tamma2014} and quantum metrology \cite{Motes2015}.

We point out that our description strongly relies on the condition of equal propagation times along any possible path at the heart of boson sampling devices. Only in this case, for a given interferometric network, the interference of all the possible multiphoton detection amplitudes characterizing the correlation function in Eq. \eqref{eqn:IntegratedGeneral} depends only on the pairwise overlap of the spectral distributions of the single photons in Eq. \eqref{eqn:g}. 

As we expected if the $N$ input photons are completely distinguishable, i.e. all their spectral distributions do not overlap, no multiphoton interference can be observed and the BSP becomes trivial.  
	If the spectra of all photons are identical, then full $N$-photon interference is observed and the probability depends on a single permanent of a random complex matrix (see Eq.~\eqref{eqn:IdenticalPhots}). This is the only situation for which the complexity of boson sampling has so far been argued \cite{aaronson2011computational}. What happens in the case of $N$ partially distinguishable photons \cite{TammaLaibacher2015,Tichy2015,Shchesnovich2015}?

In order to answer these questions it can be useful to investigate the expression of the probability in Eq. \eqref{eqn:IntegratedGeneral}. For $N$ partially indistinguishable input photons all the permanents in Eq. \eqref{eqn:IntegratedGeneral} contribute to the total probability with different weights depending on the partial overlap between the photonic spectra $\vec{\xi}_{\sources}(\omega)$ and $\vec{\xi}_{\sources '}(\omega)$ $\forall s \neq s'$. Differently, for identical photons  all the weights are equal leading to the result in Eq.~\eqref{eqn:IdenticalPhots}. Moreover, here the matrices of interest are not the random matrices $\mathcal{U}^{(\mathcal{D},\mathcal{S})}$ in Eq.~\eqref{eqn:SubU} but the interference type matrices in Eq. \eqref{A}.
The answer about the complexity of the sum of such matrices in Eq. \eqref{eqn:IntegratedGeneral} can lead to a deeper connection between      
multiboson interference and computational complexity beyond the case of identical photons in ``non-resolving'' measurements. 

We also gave a full description for the case where only $N'<N$ photons are completely indistinguishable while the rest of them are distinguishable. This leads to a detection probability in Eq. (\ref{NN'}) that, for $N'$ already of the order of a few dozens, corresponds to the weighted sum of $N!/(N'!(N-N')!)$ permanents, each one not tractable classically. What has complexity theory to say about the sum itself?
Also, this question has so far not been answered.
As we have demonstrated in \cite{TammaLaibacher2015,LaibacherTammaComplexity}, the question of complexity of sampling with non-identical photons might be easier to answer if the problem of multi-boson correlation sampling with time- and polarization-resolving detectors is considered instead of the standard boson sampling. Indeed, multi-boson correlation sampling is clearly at least as complex as standard boson sampling for all degrees of distinguishability. Moreover, in this case, the detection probabilities are only determined by a single permanent. This has allowed us to argue in \cite{LaibacherTammaComplexity} that multi-boson correlation sampling is complex even for photons that are distinguishable in the sense of Eq.~\eqref{eqn:g}.

In conclusion, this work paves the way towards a deeper understanding of the physics and the complexity of linear optical networks with input photons of arbitrary spectra.

\section*{Funding}
V.T. acknowledges the support of the German Space Agency DLR with funds provided by the Federal Ministry of Economics and Technology (BMWi) under grant no. DLR 50 WM 1556.
This work was supported by a grant from the Ministry of Science, Research and the Arts of Baden-W\"urttemberg (Az: 33-7533-30-10/19/2).

%
\end{document}